\RequirePackage{lineno}
\documentclass[twocolumn,aps,prc,showpacs,superscriptaddress,floatfix]{revtex4-2}
\usepackage{rotating}
\usepackage{epsfig}
\usepackage[colorlinks=true, allcolors=blue]{hyperref}
\usepackage{lineno}
\usepackage{amsmath}
\usepackage{graphicx}
\usepackage[caption=false]{subfig}
\usepackage{xcolor}

\usepackage{xparse,xcoffins}

\ExplSyntaxOn
\NewCoffin\imagecoffin
\NewCoffin\labelcoffin

\keys_define:nn { fqwang/label }
 {
  label   .tl_set:N = \l_fqwang_label_tl,
  labelbox .bool_set:N = \l_fqwang_label_box_bool,
  labelbox .default:n = true,
  fontsize .tl_set:N = \l_fqwang_label_size_tl,
  fontsize .initial:n = \normalsize,
  pos .choice:,
  pos/nw .code:n = \tl_set:Nn \l_fqwang_label_pos_tl { left,up },
  pos/ne .code:n = \tl_set:Nn \l_fqwang_label_pos_tl { right,up },
  pos/sw .code:n = \tl_set:Nn \l_fqwang_label_pos_tl { left,down },
  pos/se .code:n = \tl_set:Nn \l_fqwang_label_pos_tl { right,down },
  pos/nwhigh .code:n = \tl_set:Nn \l_fqwang_label_pos_tl { left,uphigh },
  pos/nehigh .code:n = \tl_set:Nn \l_fqwang_label_pos_tl { right,uphigh },
  pos/swhigh .code:n = \tl_set:Nn \l_fqwang_label_pos_tl { left,downhigh },
  pos/sehigh .code:n = \tl_set:Nn \l_fqwang_label_pos_tl { right,downhigh },
  pos/nwlow .code:n = \tl_set:Nn \l_fqwang_label_pos_tl { left,uplow },
  pos/nelow .code:n = \tl_set:Nn \l_fqwang_label_pos_tl { right,uplow },
  pos/swlow .code:n = \tl_set:Nn \l_fqwang_label_pos_tl { left,downlow },
  pos/selow .code:n = \tl_set:Nn \l_fqwang_label_pos_tl { right,downlow },
  pos/nwHigh .code:n = \tl_set:Nn \l_fqwang_label_pos_tl { left,upHigh },
  pos/neHigh .code:n = \tl_set:Nn \l_fqwang_label_pos_tl { right,upHigh },
  pos/swHigh .code:n = \tl_set:Nn \l_fqwang_label_pos_tl { left,downHigh },
  pos/seHigh .code:n = \tl_set:Nn \l_fqwang_label_pos_tl { right,downHigh },
  pos/nwLow .code:n = \tl_set:Nn \l_fqwang_label_pos_tl { left,upLow },
  pos/neLow .code:n = \tl_set:Nn \l_fqwang_label_pos_tl { right,upLow },
  pos/swLow .code:n = \tl_set:Nn \l_fqwang_label_pos_tl { left,downLow },
  pos/seLow .code:n = \tl_set:Nn \l_fqwang_label_pos_tl { right,downLow },
  pos/n .code:n = \tl_set:Nn \l_fqwang_label_pos_tl { hc,up },
  pos/w .code:n = \tl_set:Nn \l_fqwang_label_pos_tl { left,vc },
  pos/s .code:n = \tl_set:Nn \l_fqwang_label_pos_tl { hc,down },
  pos/e .code:n = \tl_set:Nn \l_fqwang_label_pos_tl { right,vc },
  pos .initial:n = nw,
  unknown .code:n   = \clist_put_right:Nx \l_fqwang_label_clist
                       { \l_keys_key_tl = \exp_not:n { #1 } }
 }

\clist_new:N \l_fqwang_label_clist
\box_new:N \l_fqwang_label_box
\box_new:N \l_fqwang_label_image_box

\NewDocumentCommand{\xincludegraphics}{O{}m}
 {
  \group_begin:
  \tl_clear:N \l_fqwang_label_tl
  \clist_clear:N \l_fqwang_label_clist
  \keys_set:nn { fqwang/label } { #1 }
  \tl_if_empty:NTF \l_fqwang_label_tl
   {
    \fqwang_includegraphics:Vn \l_fqwang_label_clist { #2 }
   }
   {
    \SetHorizontalCoffin\imagecoffin
     {
      \fqwang_includegraphics:Vn \l_fqwang_label_clist { #2 }
     }
    \SetHorizontalCoffin\labelcoffin
     {
      \raisebox{\depth}
       {
        \bool_if:NTF \l_fqwang_label_box_bool
         { \fcolorbox{white}{white}{\l_fqwang_label_size_tl\l_fqwang_label_tl} }
         { \l_fqwang_label_size_tl\l_fqwang_label_tl }
       }
     }
    \SetVerticalPole\imagecoffin{left}{36pt+\CoffinWidth\labelcoffin/2}
    \SetVerticalPole\imagecoffin{right}{\Width-36pt-\CoffinWidth\labelcoffin/2}
    \SetHorizontalPole\imagecoffin{up}{\Height-12pt-\CoffinHeight\labelcoffin/2}
    \SetHorizontalPole\imagecoffin{down}{12pt+\CoffinHeight\labelcoffin/2}
    \SetHorizontalPole\imagecoffin{uphigh}{\Height-6pt-\CoffinHeight\labelcoffin/2}
    \SetHorizontalPole\imagecoffin{downlow}{6pt+\CoffinHeight\labelcoffin/2}
    \SetHorizontalPole\imagecoffin{uplow}{\Height-18pt-\CoffinHeight\labelcoffin/2}
    \SetHorizontalPole\imagecoffin{downhigh}{18pt+\CoffinHeight\labelcoffin/2}
    \SetHorizontalPole\imagecoffin{upHigh}{\Height-0pt-\CoffinHeight\labelcoffin/2}
    \SetHorizontalPole\imagecoffin{downLow}{0pt+\CoffinHeight\labelcoffin/2}
    \SetHorizontalPole\imagecoffin{upLow}{\Height-24pt-\CoffinHeight\labelcoffin/2}
    \SetHorizontalPole\imagecoffin{downHigh}{24pt+\CoffinHeight\labelcoffin/2}
    \use:x{\JoinCoffins\imagecoffin[\l_fqwang_label_pos_tl]\labelcoffin[vc,hc]} 
    \TypesetCoffin\imagecoffin
   }
   \group_end:
 }
\NewDocumentCommand{\setlabel}{m}
 {
  \keys_set:nn { fqwang/label } { #1 }
 }
\cs_new_protected:Nn \fqwang_includegraphics:nn
 {
  \includegraphics[#1]{#2}
 }
\cs_generate_variant:Nn \fqwang_includegraphics:nn { V }

\ExplSyntaxOff

\newcommand\mean[1] {\left\langle#1\right\rangle}

\newcommand\snn     {\sqrt{s_{_{\rm NN}}}}
\newcommand\gevc    {GeV/$c$}
\newcommand\gevcc   {GeV/$c^2$}
\newcommand\bkg     {{\rm bkg}}
\newcommand\kpair   {K^+K^-}
\newcommand\kk      {{KK}}
\newcommand\ptkk    {\ensuremath{p_{\perp,\kk}}}

\newcommand\pt      {\ensuremath{p_{\perp}}}
\newcommand\bt      {\ensuremath{\beta_{\perp}}}
\newcommand\xt      {\ensuremath{x_{\perp}}}
\newcommand\betat   {\ensuremath{\beta_{\perp}}}

\newcommand\mc      {\ensuremath{{\rm{MC}}}}
\newcommand\fm      {\phi{\mathrm{\mbox{-}meson}}}
\newcommand\rzero   {\rho_{00}}
\newcommand\minv    {m_{\rm inv}}
\newcommand\cthe    {\cos^2\theta^*}
\newcommand\mcos    {\mean{\cthe}}

\begin{document}

\title{Impact of Tracking Resolutions on $\phi$-Meson Spin Alignment Measurement}

\author{C.W.~Robertson}
%\email{rober558@purdue.edu}
\affiliation{Department of Physics and Astronomy, Purdue University, West Lafayette, IN 47907}
\author{Yicheng Feng}
%\email{feng216@purdue.edu}
\affiliation{Department of Physics and Astronomy, Purdue University, West Lafayette, IN 47907}
\author{Fuqiang Wang}
%\email{fqwang@purdue.edu}
\affiliation{Department of Physics and Astronomy, Purdue University, West Lafayette, IN 47907}

\begin{abstract}
Measurements of global spin alignment of vector mesons in relativistic heavy-ion collisions can provide unique insights into spin-orbit interactions and vector meson dynamics in the Quark-Gluon Plasma (QGP) produced in those collisions. 
The global spin alignment is measured by the $00^{\rm th}$ coefficient of the spin density matrix,  $\rzero$, via the polar angle ($\theta^{*}$) of the decay-daughter momentum in the parent rest frame with respect to the direction of the orbital angular momentum of the collision.
Such measurements are affected by the angular and momentum resolutions of the reconstructed tracks in the experiment.
%In experiment, the momentum and angular resolution of a given track is always finite and therefore any tracking inaccuracy will affect vector meson spin alignment measurements. 
%The kinematics of these measurements are non-trivial, and the signal is weak (few percent), so the effect of momentum resolution is significant. 
Such effects are nontrivial because of kinematic complications caused by the boost to the parent rest frame, and could be important given that the global spin alignment signal is weak. 
In this paper, we investigate the effects of experimental tracking resolutions on measurements of the $\phi$(1020) meson $\rzero$. 
We study these effects for two methods of $\rzero$ measurements, the conventional method analyzing the $\fm$\ yield versus $\cthe$ and the invariant mass ($\minv$) method utilizing $\mcos$ versus $\minv$.
Using typical resolution values from experiments, we find that the effect of track resolution on $\rzero$ is small, well within typical measurement uncertainties.
\end{abstract}
%\pacs{25.75.-q,25.75.Ld}% PACS

\maketitle

%--------------------------------------------------------------------------------------------------
\section{Introduction}
A large orbital angular momentum (OAM) is present in non-central heavy-ion collisions~\cite{Liang:2004ph,Voloshin:2004ha}. 
This large OAM produces a vorticity field in the Quark-Gluon Plasma (QGP) which can polarize spin-1/2 quarks in the medium~\cite{Liang:2004ph,Wu:2019eyi,Becattini:2020ngo,Sheng:2019kmk}. 
These polarizations impact final state hadron production and can be measured by parity-violating weak decays of hyperons and by parity-conserving strong decays of vector mesons~\cite{Liang:2004ph}. 
A finite global spin polarization of the $\Lambda$-hyperon has indeed been observed, suggesting the presence of an ultra-strong vorticity field in the QGP~\cite{STAR:2017ckg}. 
Finite global spin alignment of $\fm$ has recently been reported by the ALICE~\cite{ALICE:2019aid} and STAR~\cite{STAR:2022fan} experiments, with opposite signs. 
Global spin alignment is measured by the angular distribution of a daughter kaon from the $\phi\rightarrow \kpair$ decay~\cite{ALICE:2019aid,STAR:2022fan},
\begin{equation}
    \frac{dN}{d\cos{\theta^{*}}} \propto (1-\rzero) + (3\rzero -1)\cos^{2}{\theta^{*}}\,,
    \label{eq:decay}
\end{equation}
where $\theta^*$ is the polar angle of the  kaon's momentum vector in the $\fm$ rest frame with respect to the global OAM in the lab frame.
The parameter $\rzero$ is the $00^{\rm th}$ coefficient of the spin density matrix. A uniform angular distribution gives $\rzero=1/3$. The deviation of $\rzero$ from 1/3 indicates a finite spin alignment. 
Since all known physics mechanisms yield only negligible $\rzero-1/3 \sim 10^{-4}$, the relatively large values of $\sim 1\%$ measured of $\rzero-1/3$ may suggest novel physics mechanisms such as fluctuations of the strong color field~\cite{Sheng:2019kmk,Sheng:2022wsy}.

In experiments, charged-particle trajectories are reconstructed from hit information recorded in tracking devices. Measurement resolutions can cause inaccuracies in the reconstructed track parameters.
Moreover, because of small-angle multipole scattering of charged particles in detector materials, the particle trajectories will derail, the magnitude of which being dependent on the amount of detector materials they traverse.
These physics processes will cause inaccuracies in the reconstructed particle kinematics even with a perfect detector.
Since $\rzero$ is measured through the angular distribution of the decay daughters, the angular resolution of the reconstructed tracks is important. 
Because $\theta^*$ is measured in the parent rest frame, both angular and momentum resolutions can impact the Lorentz boost, which can unpredictably affect the measurement. 

In this article, we investigate the effects of angular and momentum resolutions on $\rzero$ measurements using toy-model simulations. 

%================================================================================================
\section{Simulation and Analysis Details\label{sec:details}}

\subsection{Track Angular and Momentum Resolutions}
The two-dimensional angular spread due to multiple small-angle scattering of singly-charged particles in detector materials is given by~\cite{Workman:2022ynf}
\begin{equation}
    \sigma_{\rm ms} \approx \frac{13.6 {\rm MeV}/c}{\beta p}\sqrt{\frac{x}{X_0}}
    \approx \frac{13.6 {\rm MeV}/c}{\beta\sqrt{p\pt}}\sqrt{\frac{\xt}{X_0}}\,,
    \label{eq:ms}
\end{equation}
where $p$ and $\beta$ are particle momentum magnitude and velocity in units of the speed of light, and $x/X_0$ is the amount of materials the particle traverses in units of the radiation length. 
It is convenient to use the transverse material length $x_\perp/X_0$, independent of the particle's polar angle.

In addition, tracking accuracy is affected by the detector position resolution.
The angular resolution of the track trajectory is given by the position resolution over the typical track length, which is more or less fixed by the size of the detector. Thus, the detector position resolution impacts the angular resolution in an approximately uniform way. 
%Thus, the detector position resolution contributes an approximately constant to the angular resolution.

Experimentally, one typically measures $\pt$ and pseudorapidity $\eta$. Within a given $\eta$ range close to midrapidity, $p$ is approximately proportional to $\pt$ on average. The average effect of multiple scattering is therefore inversely proportional to $\bt\pt$.
Thus, the standard deviations in the polar ($\theta$) and azimuthal ($\phi$) angles can be parameterized with the following function: 
\begin{equation}
    \sigma_{\delta\theta,\delta\phi} = \sqrt{ \left(\frac{A_{\theta,\phi}}{\betat\pt}\right)^2 + B_{\theta,\phi}^2}\,,
    \label{eq:angle}
\end{equation}
where $\betat\equiv\pt/\sqrt{\pt^2+m^2}$ ($m$ is the particle rest mass). 
Note that we have used $\bt$ in place of $\beta$, which are not strictly proportional to each other. They are equal for midrapidity particles. However, in practice, Eq.~(\ref{eq:angle}) is a good parameterization for angular resolutions. While $A_\theta\approx A_\phi\approx\sigma_{\rm ms}$, $B_\theta$ and $B_\phi$ can differ depending on the position resolutions (granularity) in the polar and azimuthal directions.

The inverse of the particle $\pt$ is determined by the track curvature in the transverse plane ($\rho_\perp$), $1/\pt\approx\rho_\perp/(0.3B)$ where $B$ is the detector magnetic field.
Track curvature is equal to the angular bend of the track over the detector range ($L$), so the precision of $\pt$ and $\phi$ are related by $\delta\rho_\perp\sim\delta\phi/L$. The relative momentum resolution is therefore given by the azimuthal resolution multiplied by $\pt$,
\begin{equation}
    \sigma_{\delta\pt/\pt} = \frac{1}{0.3BL} \sqrt{\left(\frac{A_\phi}{\betat}\right)^2 + (B_\phi\pt)^2}\,,
    \label{eq:pt}
\end{equation}
where $B, L$ and $\pt$ are in the units of Tesla, meter, and \gevc, respectively. 

Typically in experiments, the angular and momentum resolutions are obtained from GEANT simulation of {\em Monte Carlo} (MC) tracks going through the detector materials~\cite{STAR:2008med}. 
To illustrate the effects of tracking resolution on spin alignment measurements, we consider the STAR Time Projection Chamber (TPC)~\cite{Anderson:2003ur} as a typical detector. The amount of STAR TPC transverse material is approximately $\xt/X_0\approx1\%$. This would give $\sigma_{\rm ms}\sim 0.00136$ for a $\pt=1$~\gevc\ light hadron at midrapidity. 
Typical position granularity gives an angular resolution on the order of per mille for straight-line tracks (high $\pt$).
We therefore use the following parameters in our subsequent simulations:
\begin{subequations}\label{eq:res}
    %A_\theta=A_\phi=0.0015~{\rm GeV}/c, \;\;\; B_\theta=B_\phi=0.0010\,.
\begin{align}
    A_\theta=A_\phi=& 0.0015~{\rm GeV}/c\,, \\
    B_\theta=B_\phi=& 0.0010\,.
\end{align}
\end{subequations}
We take the typical $B=0.5$~Tesla and $L=1$~m (like for the STAR detector~\cite{STAR:2002eio}) for momentum resolution.
\begin{figure}[hbt]
    \xincludegraphics[width=0.45\textwidth,pos=nwLow,label=\hspace*{1.2cm}a)]{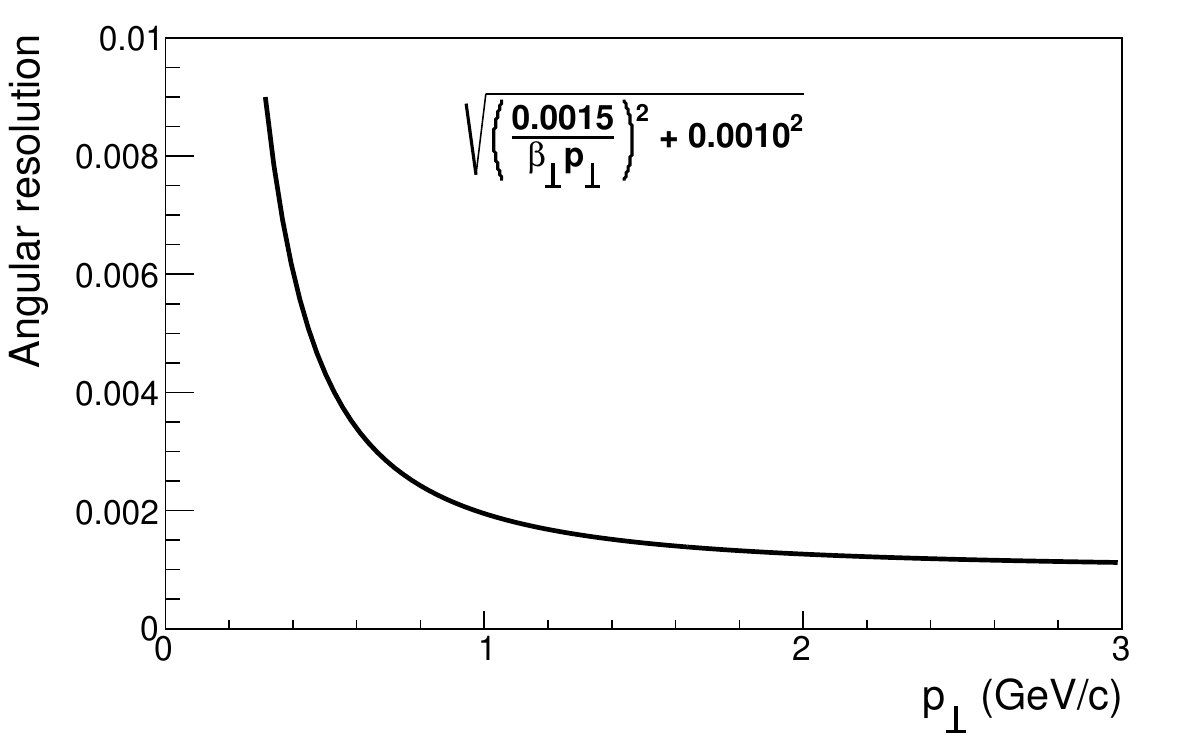}
    \xincludegraphics[width=0.45\textwidth,pos=nwLow,label=\hspace*{1.2cm}b)]{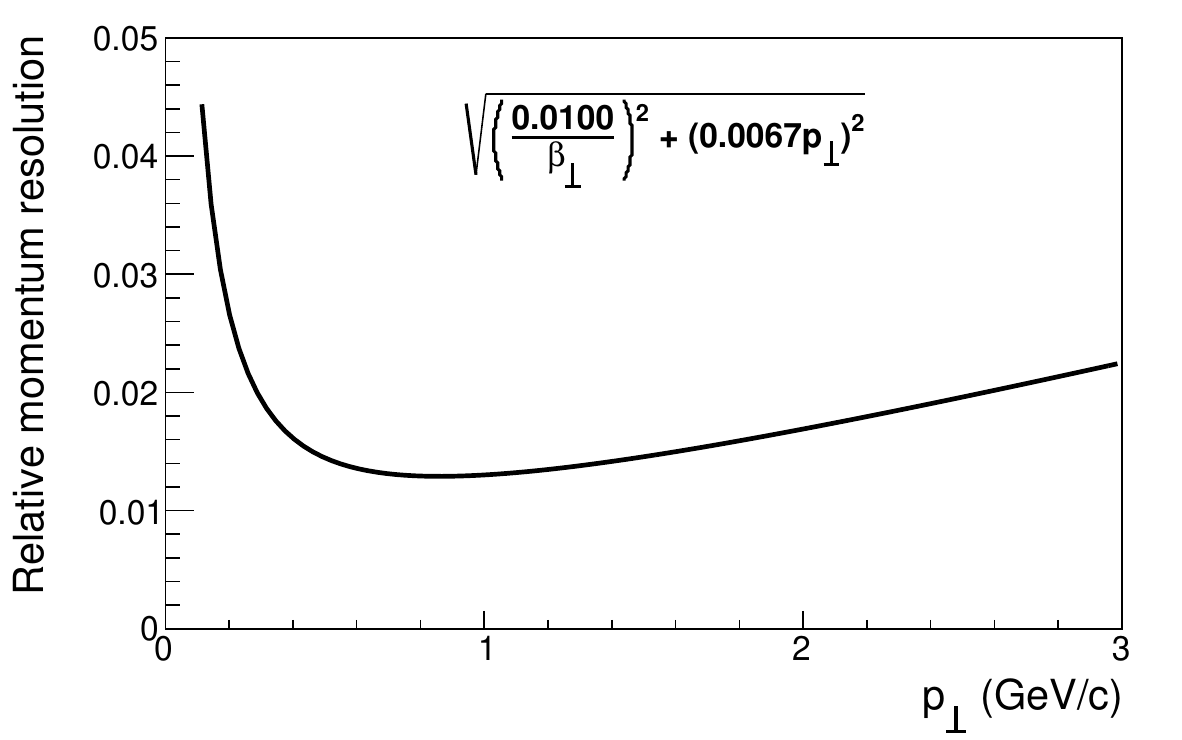}
    \caption{The parameterized (a) angular and (b) momentum resolutions of Eq.~(\ref{eq:angle}) and  Eq.~(\ref{eq:pt}), respectively, used in our toy-model simulation.}
    \label{fig:res}
\end{figure}

%--------------------------------------------------------------------------------------------------
\subsection{Setup of Toy-Model Simulations}
We generate $\phi$ mesons and kaons to mimic real data events, 
according to the measured yields, $\pt$ spectra, and $v_2(\pt)$ of $\fm$~\cite{STAR:2007mum} and charged kaons~\cite{STAR:2008med} in Au+Au collisions at $\snn=200$~GeV, respectively.
%\gray{We generate charged kaons and $\phi$ mesons with uniform $\eta$ distributions within $|\eta| <1.0$.}

The kaon $\pt$ is sampled in $ 0 < \pt < 10$~\gevc\ according to the Boltzmann distribution with a temperature given by the kinetic freeze-out temperature and then boosted by a common flow velocity. The kinetic freeze-out temperature and flow velocity are taken from Ref.~\cite{STAR:2008med} according to the measurements in 40-50\% centrality Au+Au collisions at 200 GeV. The kaon $v_2$ is taken to be $v_2= 0.06\times\pt/$\gevc, up to $\pt=2$~\gevc\ and constant above this $\pt$ value, according to the charged hadron measurement~\cite{STAR:2004jwm}. 

For $\phi$ mesons, $\pt$ is sampled within $ 1.2 < \pt < 5.4$~\gevc\ according to a fit to the $\pt$ spectrum measured in 40-50\% centrality Au+Au collisions~\cite{STAR:2007mum}. The $\fm$ $v_2$ is taken to be $v_2=0.064 \times \pt/$\gevc, from fit to $\fm$\ 200 GeV data in the 10-40\% centrality measured in \cite{STAR:2007mum} up to $\pt =3$~\gevc\ and constant above this $\pt$ value. 

The average multiplicity density for $\fm$s is taken to be $dN_\phi/dy = 1.44$ as measured in the 40- 50\% centrality~\cite{Abelev:2008aa}. 
The kaon multiplicity density $dN_{K^+}/dy=dN_{K^-}/dy=9.35$ is used and taken to be the average of the measured $K^+$ and $K^-$ multiplicities~\cite{STAR:2008med}. The event-by-event multiplicities are generated by Poisson distributions with twice these averages. 
%\gray{The $\fm$'s rapidity is uniformly sampled within $|y|<1$, and the kaon pseudorapidity is uniformly sampled within $|\eta|<1$; Pseudorapidity is used for kaons by choice to conform with data analysis~\cite{STAR:2022fan}.}
For simplicity, both the $\fm$s and the kaons are generated with uniform distributions in pseudorapidity within $|\eta|<1$.

We decay the $\fm$s into kaons according to the angular distribution of Eq.~(\ref{eq:decay}) with an input $\rzero^\mc$ value. 
We smear the direction of the momentum vectors of the kaons (both primordial and decay) by Gaussian distributions, with standard deviations according to the parameterization in Eq.~(\ref{eq:angle}). 
We smear the $\pt$ of the kaons using a Gaussian distribution with standard deviation parameterized according to Eq.~(\ref{eq:pt}). 

%--------------------------------------------------------------------------------------------------
\subsection{Analysis Methods of Spin Alignment\label{sec:method}}
We require the kaon pairs ($KK$) included in the analysis to have kinematics of  $1.2<\ptkk<5.4$~\gevc\ and $|y_\kk| < 1.0$ (where the $\fm$ mass is used in calculating $y_\kk$) as in the STAR results~\cite{STAR:2022fan}.

We examine two methods in analyzing $\rzero$: the yield method and the invariant mass ($\minv$) method. The yield method is conventional~\cite{ALICE:2019aid,STAR:2022fan}. It analyzes $\fm$\ yield vs.~$\cos{\theta^{*}}$ and extracts $\rzero$ from a fit over $\cos{\theta^{*}}$ bins, as in Eq.~(\ref{eq:decay}). 
To extract the $\fm$ yield, a mixed-event $\minv$ distribution of $\kpair$ pairs is first subtracted from that of real events; this removes the majority of the combinatorial background. The  resulting $\minv$ distribution is then fit with a Breit-Wigner (BW) function together with a certain function to model the residual background,
\begin{equation}
    %\frac{dN_{\kpair}}{d\minv} - \left(\frac{dN_{\kpair}}{d\minv}\right)_{\rm mix} = 
    \frac{dN_{\kpair}}{d\minv} = \frac{A \Gamma}{(\minv-m_\phi)^2+(\Gamma/2)^2} + \left(\frac{dN_{\kpair}}{d\minv}\right)_{\bkg}\,,
    \label{eq:yield}
\end{equation}
where the BW amplitude $A$ and width $\Gamma$ are fit parameters, and $m_\phi$ is the mass of the $\fm$ which can be treated as a fit parameter or fixed (in our analysis it is treated as a fit parameter).
In the published STAR data~\cite{STAR:2022fan} the residual background was assumed to be a first-order polynomial function. The $\fm$ yield is obtained from the fit BW function. 
If the combinatorial background was precisely known and the $\fm$\ signal shape is precisely BW, then the $\fm$\ yield can be reliably obtained in each $\cos\theta^*$ bin and the $\rzero$ be reliably extracted. A major systematic uncertainty in such analyses is the background modeling in each $\cos{\theta^{*}}$ bin, and several other functional forms are assessed for the systematic uncertainty. Another systematic uncertainty comes from the BW assumption for the signal shape. 
%This systematic uncertainty is not straightforward to assess and is not exhaustively examined in the published data. We will show in this article that this part of the systematics may be important.
This systematic uncertainty is not straightforward to assess, but may be important.

The invariant mass method is newly developed. The idea is not new; similar methods have been used in hyperon polarization analysis and in anisotropic flow analysis of resonances. First, the $\cos{\theta^{*}}$-inclusive $\minv$ distribution is fit by BW and a background function, typically polynomial, and the $\fm$\ signal-to-background ratio, $r(\minv)$, is obtained from the fit~\footnote{One can also subtract the mixed-event background first, and then do a fit with BW and a residual background to obtain the signal-to-noise ratio with the subtracted mixed-event background added back. Since the purpose of this fit is to obtain the signal-to-noise ratio of $\fm$, not the yield vs.~$\cos\theta^*$ as in the yield method, the different ways of fit do not cause significant systematics}. 
The $r(\minv)$ is  then used to fit a profile of the $\mcos$ vs.~$\minv$ by
\begin{equation}
    \mcos = \frac{r(\minv) (5\rzero - 1)/2 + \mcos_{\bkg}}{r(\minv) + 1}\,,
    \label{eq:mcos}
\end{equation}
to extract the spin alignment $\rzero$ parameter.
In Eq.~(\ref{eq:mcos}) we have converted $\mcos$ of the $\phi$-meson decay Eq.~(\ref{eq:decay}) into $\rzero$, $\mcos=\frac{5}{2}(\rzero - \frac{1}{3}) + \frac{1}{3}$, so $\rzero$ can be directly extracted from the fit.
This method is expected to be less vulnerable than the yield method because it does not have to fit the signal and background of the $\minv$ distributions of several $\cos\theta^*$ bins, but only the inclusive $\minv$ distribution and the fit of $\mcos$ vs.~$\minv$. 
The latter fit involves modeling the background $\mcos_\bkg$  as a function of $\minv$, which depends on the kinematics of the background kaons. 
% In the vicinity of the $\fm$ mass, the $\mcos_\bkg$ can be approximated by a linear function in $\minv$,
% \begin{equation}
%     \mcos_\bkg \propto \minv - 2m_K\,,
%     \label{eq:mcos_bkg}
% \end{equation}
% where $\mcos_\bkg=0$ by definition at the threshold of twice kaon mass.
We use a first-order polynomial in $\minv$ for the background $\mcos_\bkg$.

The details of these two analysis methods will become clearer along the way when we present the results of our study. 
Angular and momentum resolutions can affect both methods in non-trivial ways, and this will also become clearer when we discuss their effects on the results. 

%================================================================================================
\section{Simulation Results}

\subsection{Pure $\fm$ Case\label{sec:pure}}
\begin{figure}[hbt]
    \xincludegraphics[width=0.5\textwidth,pos=nwLow,label=\hspace*{4mm}a)]{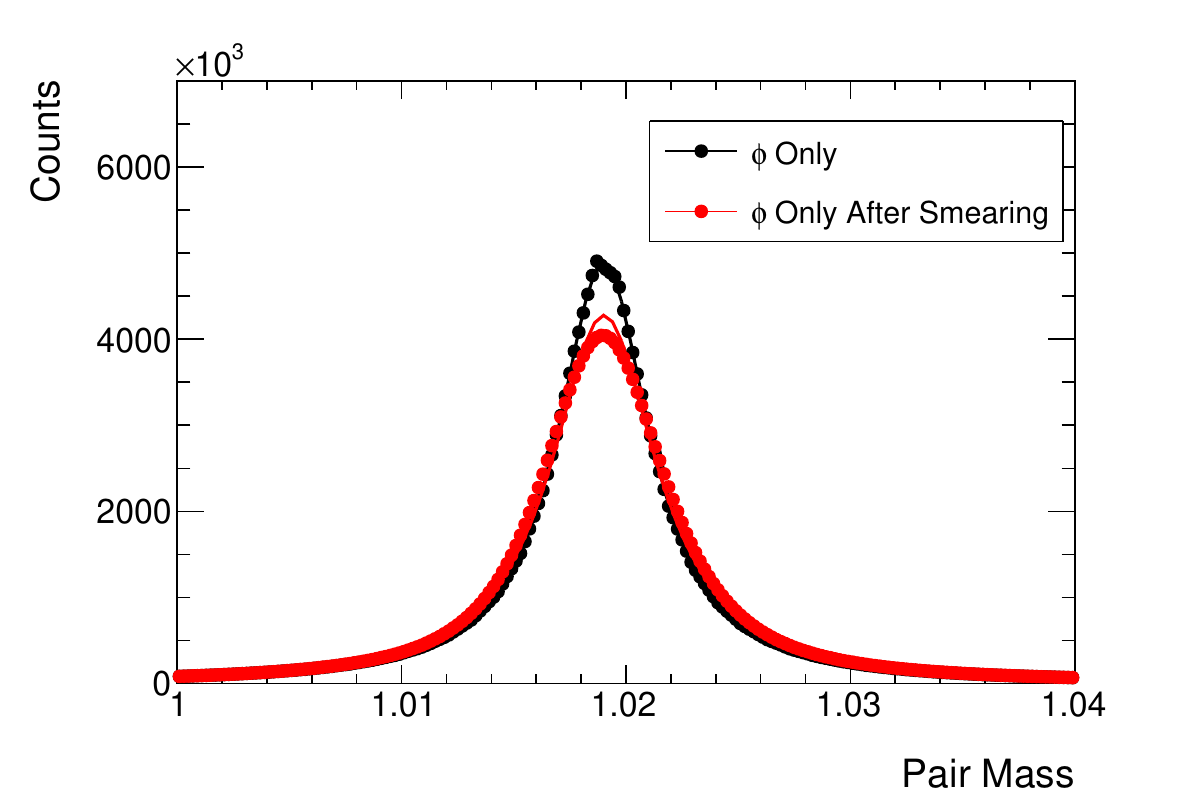}\hfill
    \xincludegraphics[width=0.5\textwidth,pos=nwLow,label=\hspace*{4mm}b)]{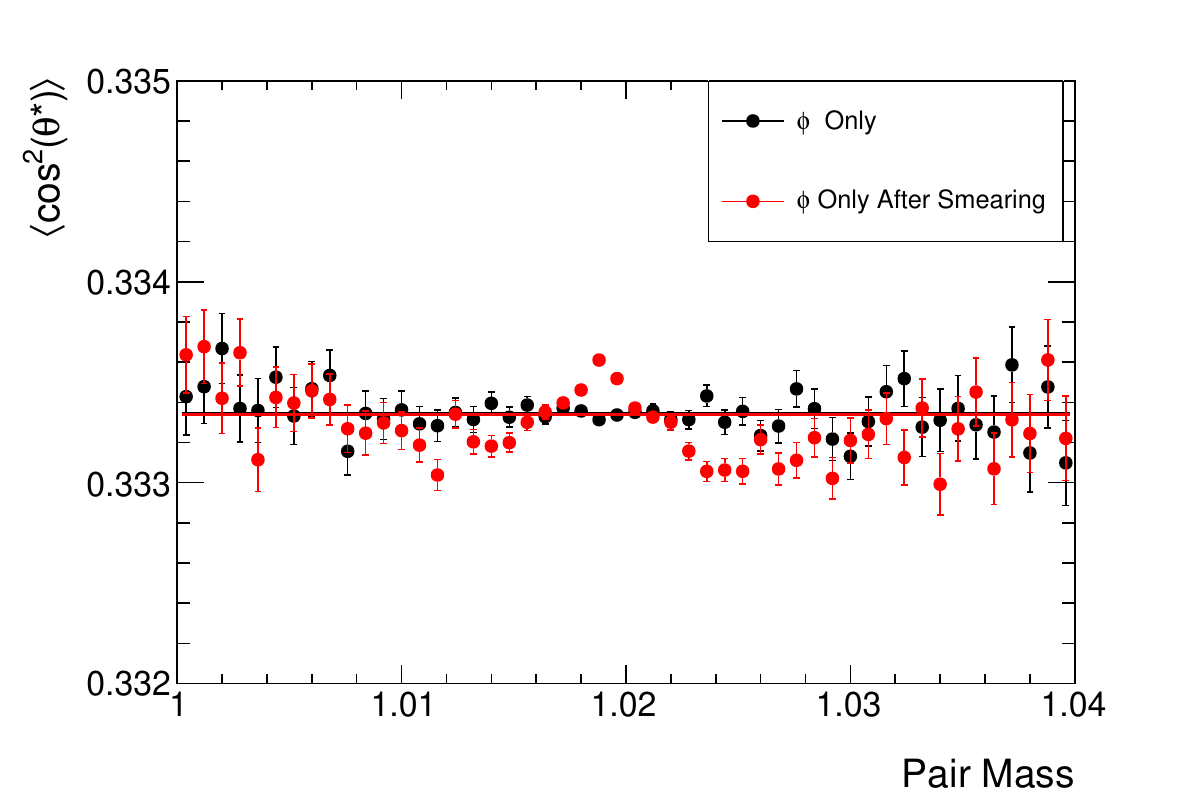}
    \caption{(Color online) Resolution effects on true $\fm$s in the invariant mass method.
    (a) $\fm$ $\minv$ distributions, and (b) $\mcos$ as a function $\minv$, before (black markers) and after (red markers) angular and momentum smearing by Eqs.~(\ref{eq:angle}--\ref{eq:res}). 
    Only those real $\fm$s (no combinatorial $\kpair$ pairs) are used in these plots and the input $\rzero = 1/3$. 
    The BW fits by Eq.~(\ref{eq:yield}) (with no background) are superimposed on (a); The horizontal lines in (b) are constant fit to the data points of corresponding color.}
    \label{fig:PhiOnly}
\end{figure}

We first examine the impact of track resolutions on the reconstructed $\mcos$ of the real $\fm$s (without including combinatorial background) as a function of $\minv$. The input of $\fm$ is given as $\rzero = 1/3$ (that is, without spin alignment). Figure~\ref{fig:PhiOnly}(a) shows the $\minv$ distributions of the $\fm$s before (black) and after (red) angular and momentum smearing. 
The curves are BW fits. 
As expected, the smeared case has a lower amplitude, larger width, and worse $\chi^2$ goodness of fit for a pure BW signal, relative to the one before smearing. In other words, the signal is no longer a BW function, and this will be discussed further below. Figure~\ref{fig:PhiOnly}(b) is the $\fm$\ $\mcos$ vs.~$\minv$ for the unsmeared (black) and smeared case (red). Without smearing, we see roughly uniform $\mcos$ vs.~$\minv$; with smearing, we see distortion near the $\fm$\ mass--there is a considerable peak at $m_\phi$ and dips at lower and higher $\minv$. 
An example constant fit is shown in Fig.~\ref{fig:PhiOnly}(b) demonstrating that the weighted average $\mcos$ over the $\fm$ mass region recovers the input $\rzero = 1/3$, or $\mcos=1/3$, despite the momentum and angular smearing.

\begin{figure}
    \includegraphics[width=0.5\textwidth]{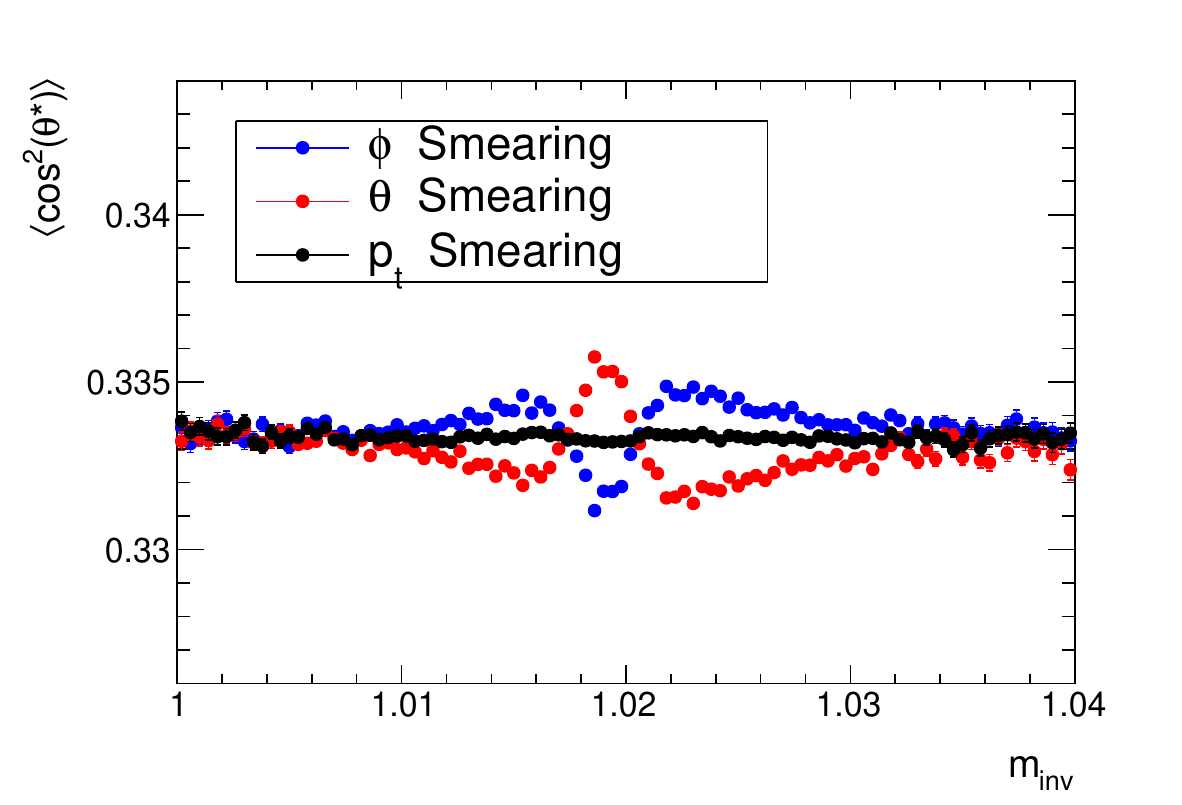}
    \caption{(Color online) Effects of smearing for $\theta$ (red markers), $\phi$ (blue markers), and $\pt$ (black markers), separately, on $\mcos$ for true $\fm$s. The effect of momentum smearing is insignificant compared to those of angular smearing. Equal $\theta$- and $\phi$-angle resolutions are used as in Eq.~(\ref{eq:res}).}
    \label{fig:PhiOnly2}
\end{figure}

Figure~\ref{fig:PhiOnly2} shows the individual effects of the $\theta$ and $\phi$ resolutions and the momentum resolution. The angular smearing has a much larger effect than the smearing in transverse momentum. 
A $\mcos$ peaked region means that the large $\cthe$ bins have more $\kk$ pairs than its fair share, and the dipped region means the opposite. 
Therefore, the center-dipped and side-peaked $\mcos$ vs.~$\minv$ distribution from the $\phi$-angle smearing implies that the $\minv$ distributions of the $\fm$s are broadened from small to large $\cthe$ bins. The opposite behavior from the $\theta$-angle smearing implies that the $\fm$ $\minv$ distributions are narrowed from small to large $\cthe$ bins.
%{These opposite behaviors are presumably because the smearing in $\theta$ and $\phi$ has a non-trivial effect on the individual transformations from lab frame to parent rest frame.}
These opposite behaviors are presumably due to non-trivial interplays of the different Jacobians involving the $\theta$ and $\phi$ angles and the kinematic boost to the pair rest frame.
Note that we have used the same resolution parameters as in Eq.~(\ref{eq:res}) for the smearing of the angles $\theta$ and $\phi$.
%The effects of polar and azimuthal angle smearing are opposite and largely cancel each other out. 
The opposite effects of polar- and azimuthal-angle smearing largely cancel each other out. 
Asymmetric polar- and azimuthal-angle resolutions would make the overall effect larger.

%--------------------------------------------------------------------------------------------------
\subsection{$\fm$ with Combinatorial Background\label{sec:real}}
We now perform a ``data-like" analysis with the $\minv$ method and the yield method with input $\rzero = 1/3$. 

\begin{figure}
    \xincludegraphics[width=0.5\textwidth,pos=nwLow,label=\hspace*{5mm}a)]{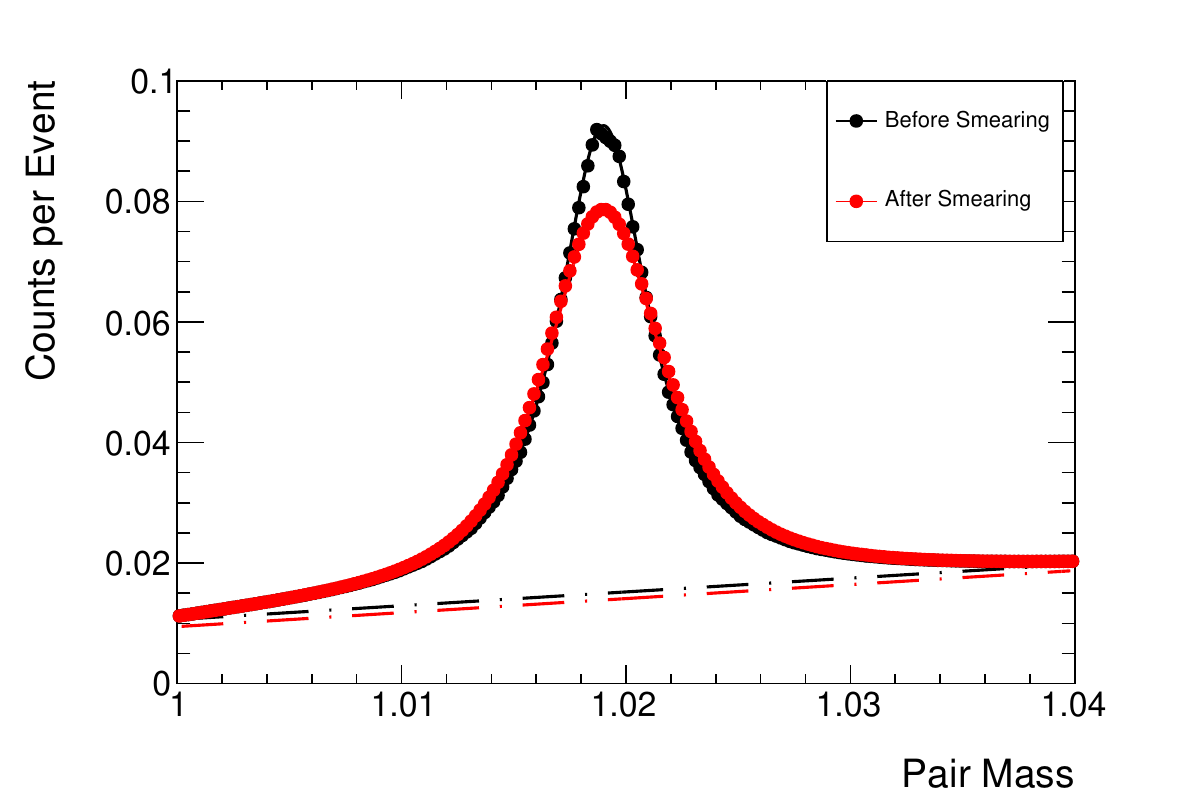}\hfill
    \xincludegraphics[width=0.5\textwidth,pos=nwLow,label=\hspace*{5mm}b)]{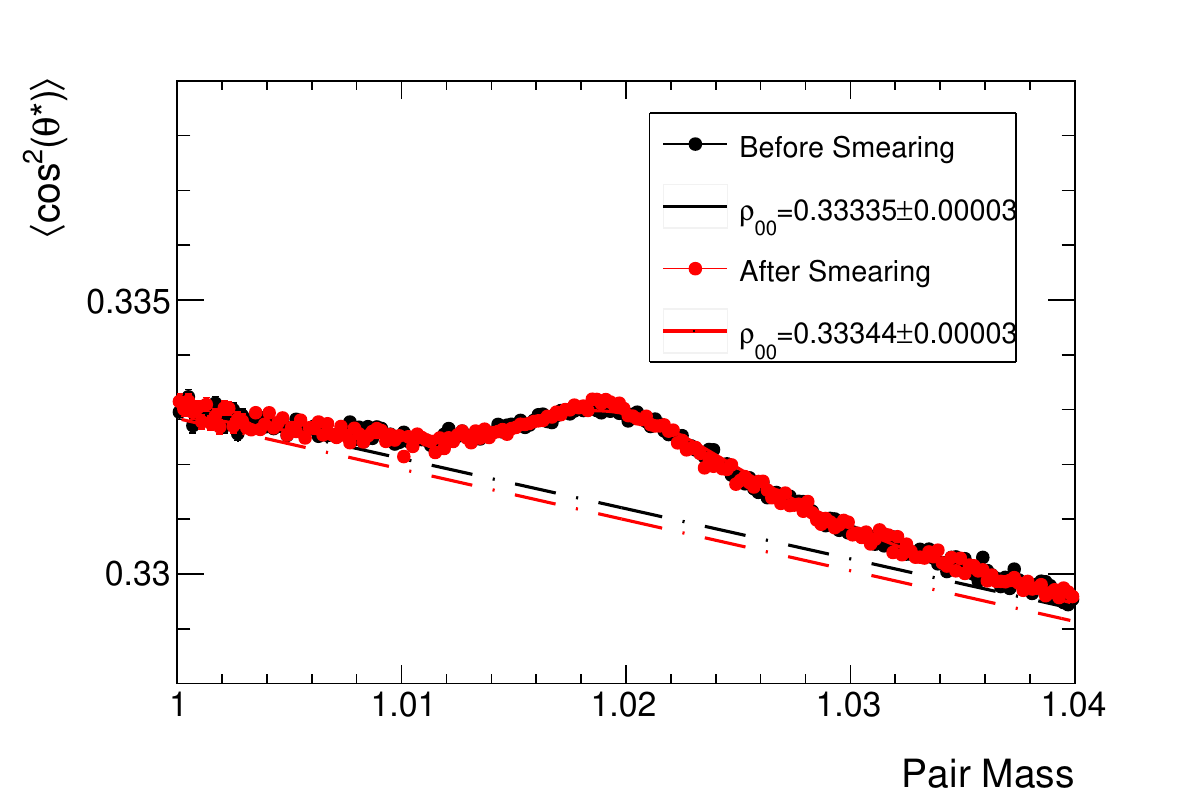}\hfill
    \caption{(Color online) Resolution effects in the invariant mass method.
    (a) $\kpair$ $\minv$ distributions, and (b) $\mcos$ as a function $\minv$, before (black markers) and after (red markers) angular and momentum smearing by Eqs.~(\ref{eq:angle}--\ref{eq:res}). 
    The kinematics of the $\fm$s and the background kaons, including $v_2$, are taken from Refs.~\cite{Mohanty:2009tz,STAR:2008med} according to the 40--50\% centrality Au+Au collisions.
    The input $\fm$ $\rzero = 1/3$.
    The BW and linear background fits by Eq.~(\ref{eq:yield}) are superimposed on (a); 
    The fits in (b) are by Eq.~(\ref{eq:mcos}) with a linear background. %; the fit to the smeared data has poor quality because the $\rzero$ is no longer independent of $\minv$.
    }
    \label{fig:Inv}
\end{figure}

Figure~\ref{fig:Inv} shows the $\kpair$ pair $\minv$ distribution in the upper panel and the $\mcos$ vs.~$\minv$ in the lower panel, similar to Fig.~\ref{fig:PhiOnly} but including a combinatorial background. The unsmeared data are shown in black, and the smeared data are shown in red. We see that angular and momentum smearing broadens the decay mass distribution in Fig.~\ref{fig:Inv}(a). %and creates distortion in the $\mcos$ vs.~$\minv$ plot.
Unlike the uniform distribution of $\mcos$ for the unsmeared pure $\fm$ case in Fig.~\ref{fig:PhiOnly}(b), the $\mcos$ distribution for the realistic case in Fig.~\ref{fig:Inv}(b) before smearing is peaked atop a decreasing pedestal. The pedestal is from the background $\kk$ pairs and their $\mcos$ is not 1/3, dictated by the single kaon kinematic distributions. 
The $\fm$s with $\rzero=1/3$, therefore, appear as a peak in the plot in Fig.~\ref{fig:Inv}(b).
The $\mcos$ vs.~$\minv$ for the smeared case is similar, with subtle differences from the unsmeared case.
For illustration, we fit the $\minv$ distributions in Fig.~\ref{fig:Inv}(a) with a BW function for the $\fm$ signal and a linear function for the combinatorial background; We obtain the signal-to-background ratio from the fit. We then fit the $\mcos$ vs.~$\minv$ plot in Fig.~\ref{fig:Inv}(b) with Eq.~(\ref{eq:mcos}) with the background function also being linear in $\minv$. We choose a linear function for the background in both fits, which is a good approximation within our limited fit range of $1<\minv<1.04$~\gevcc. 
One could open up the fit range and use more elaborate functions to study background systematics; however, this is outside the scope of the present study. 

\begin{figure}
    \includegraphics[width=\linewidth]{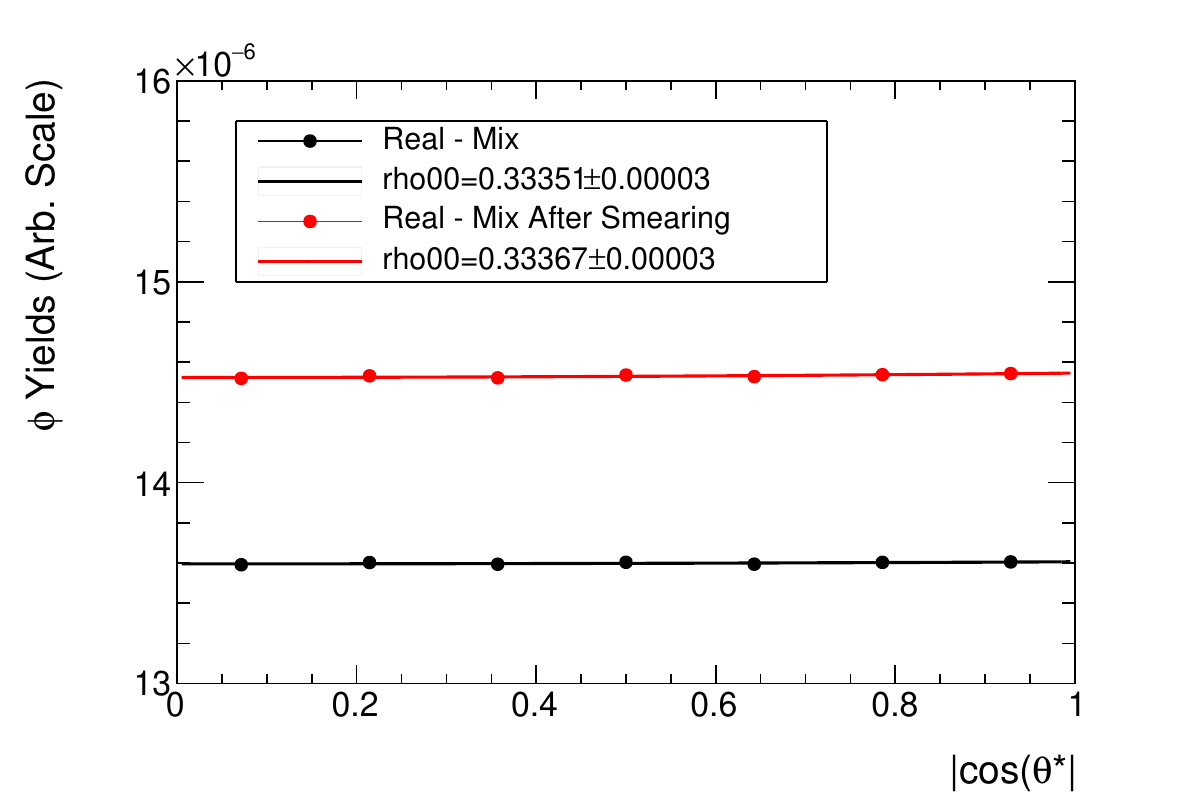}
    \caption{(Color online) Yield vs.~$|\cos\theta^*|$ bin. The $\fm$ yield is obtained from fitting to the mixed-event subtracted $\minv$ distributions by Eq.~(\ref{eq:yield}) with a linear residual background; the BW width is fixed to that obtained from the fit to the $\cos\theta^*$-inclusive data for demonstration purpose. Simulated data before and after tracking smearing are shown in black and red markers, respectively.}
    \label{fig:yield}
\end{figure}

Next, we perform an example yield method $\rzero$ extraction. For each $\cos\theta^*$ bin, we take the real event $\minv$ distribution of $\kpair$ pairs and subtract a ``mixed event" background. We then fit the resulting $\minv$ distribution with a BW signal and a linear residual background, Eq.~(\ref{eq:yield}), to extract the yield in each $\cos\theta^*$ bin. In this case, we use a fixed BW width $\Gamma$ in each bin, where this width is obtained from fitting the $\cos\theta^*$-inclusive pair distribution by Eq.~(\ref{eq:yield}). 
Alternatively, the BW width can be allowed to vary from bin to bin, which we examine later. Regardless of how the yields are extracted $\rzero$ is obtained from a fit to these yields using Eq.~(\ref{eq:decay}). Figure~\ref{fig:yield} shows the extracted $\fm$ yield as a function of $|\cos\theta^*|$ for the unsmeared (black) and smeared (red) data, superimposed by the fit curves of Eq.~(\ref{eq:decay}).
The unsmeared data are consistent with $\rzero=1/3$, while the smeared data show a slight increase in the yield with $|\cos\theta^*|$, thus a $\rzero$ larger than 1/3.

Our toy-model simulations demonstrate an effect of track resolutions on $\rzero$. 
Such an effect can be experimentally verified by comparing tracks of differing qualities. 
For example, in the STAR experiment, tracks are reconstructed from TPC hits (called global tracks), and then refitted by including the primary vertex position (called primary tracks). 
Primary tracks are more precise than global tracks; the differences in their track parameters can be considered as reflections of tracking resolutions.
It would be interesting to compare the $\rzero$ values measured with global tracks and with primary tracks. 

%--------------------------------------------------------------------------------------------------
\subsection{Strategy to Extract $\fm$ $\rzero$}
As we have seen in the previous subsections, the angular and momentum smearing can distort the $\mcos$ vs.~$\minv$ shape in non-trivial ways. 
%The smearing produces a dip in the mean $\mcos$ vs.~$\minv$ at the $\fm$  mass and two peaks on the sides of the $\fm$ mass. 
These complicated behaviors are presumably caused by the interplay of many factors, each of which is affected by resolution smearing. These factors include the $\minv$ calculated from smeared $\kpair$ pairs, the boost to the rest frame of the smeared pair, and the fact that $\theta^*$ is defined as the angle of a daughter in the pair rest frame with respect to the OAM direction in the lab frame (perpendicular to the reaction plane).

As mentioned above, the $\mcos$ peak at the $\fm$ central mass value, such as that in Fig.~\ref{fig:PhiOnly}(b), means that relatively more particles are found at the $\fm$ mass position in large $\cthe$ bins. This in turn means that the large $\cthe$ bins will have a narrower $\minv$ peak than the small $\cthe$ bins. In other words, the $\minv$ peak width decreases with increasing $\cthe$. 
This shuffling of $\kpair$ pairs across $\cthe$ bins and across $\minv$ bins also renders that the signal shape is no longer BW. 
The extraction of $\rzero$ with the assumed BW signal is therefore questionable. 
Furthermore, the width is often fixed across all $\cthe$ bins in order to have better controlled systematics. However, resolution smearing narrows the mass peak with increasing $\cthe$; This makes the extracted yields more questionable.

\begin{figure}
    \includegraphics[width=\linewidth]{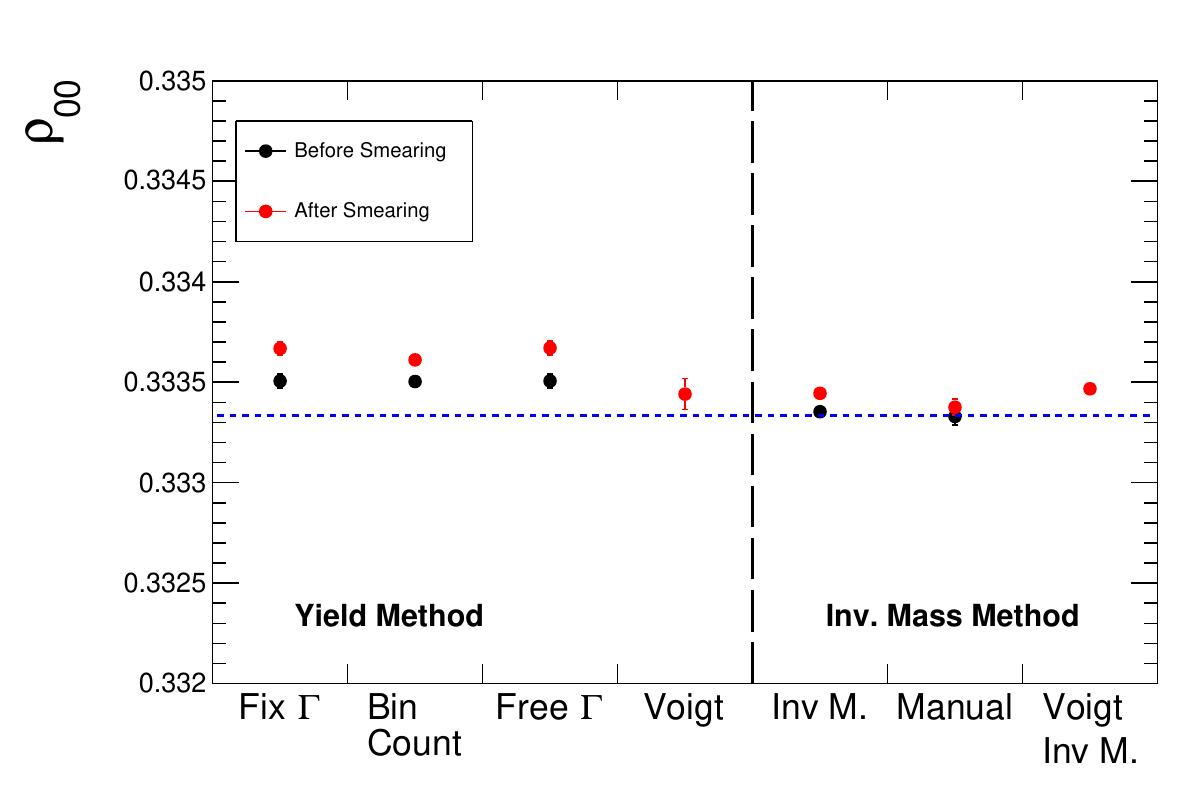}
    \caption{(Color online) Extracted $\rzero$ by various methods from simulation data before (black markers) and after (red markers) track smearing. 
    The methods used are from left to right: the yield method by Breit-Wigner fit of Eq.~(\ref{eq:yield}) with fixed width (fit yield and bin counting within twice widths both sides), by Breit-Wigner fit with free width, and by Voigt fit with free width; the invariant mass method by Breit-Wigner fit of Eq.~(\ref{eq:yield}) and then a fit to $\mcos$ vs.~$\minv$ by Eq.~(\ref{eq:mcos}), by manual calculation (bin counting) subtracting the fitted backgrounds, and by Voigt fit and then the $\mcos$ fit of Eq.~(\ref{eq:mcos}). See text for details.}
    \label{fig:rho}
\end{figure}

Figure~\ref{fig:rho} shows the extracted $\rzero$ in Section~\ref{sec:real} assuming the BW function for the $\fm$ signal (with both fixed and varying widths) and a linear function for the residual combinatorial background. 
We also show the $\rzero$ value obtained with yields from bin counting within a mass range of $\pm 2\Gamma$ about the peak mass from the fixed-width fit. The extracted $\rzero$ without resolution smearing (black points) is slightly above the input value of $1/3$, presumably because of systematics in the yield extraction, e.g., the imperfect linear background assumption.  The extracted $\rzero$ values with resolution smearing (red points) differ from those without, indicating the effect of track resolutions on $\rzero$. However, the effect is small, less than 0.0005, well within the typical experimental uncertainties of the order of 0.001~\cite{STAR:2022fan}. We note, as mentioned in Sect.~\ref{sec:pure}, that we used the same angular resolution for polar and azimuthal angles and unequal resolutions between them would likely increase the effect of resolution smearing on $\rzero$. 

We may improve the yield method in several ways. For example, after momentum smearing, the signal peak is no longer BW and may be better modeled by the Voigt function (a convolution of BW with Gaussian mass resolution). The slight difference in the extraction of the yield from a Voigt function compared to the BW across $\cos{\theta^{*}}$ bins can have an effect on the extracted $\rzero$, as shown in Fig.~\ref{fig:rho}.

In the invariant mass method, we fit the $\minv$ histogram within  $1.0<\minv<1.04$~\gevcc\ using BW and a linear background function to extract the signal-to-noise ratio, and then fit the $\mcos$ vs.~$\minv$ using the signal-to-noise ratio and a linear $\mcos_{\bkg}$ function. 
The results are shown in Fig.~\ref{fig:rho}. The effect of track smearing in the invariant mass method is similar to that in the yield method.
We can improve the extraction of the signal-to-noise ratio by fitting the $\minv$ histogram using a Voigt signal function. The $\rzero$ using the new signal-to-noise ratio is also shown in Fig.~\ref{fig:rho}; no significant change is observed.

Lastly, due to the shuffle of pairs across $\cos\theta^*$ bins and across $\minv$ bins, the $\rzero$ quantity is no longer a constant over $\minv$ for the signal $\fm$s. The $\rzero$ value, treated as a fit parameter and extracted from fitting to Eq.~(\ref{eq:mcos}), may be biased. To circumvent this, we calculate $\rzero$ ``by hand,'' analogous to bin counting.
To do this, we multiply the pair count by $\mcos$ for each bin of $\minv$ within [1.01,1.03]~\gevcc, subtract the corresponding background quantity, and divide by the total number of signal pairs. 
The background pair count and the background $\mcos_\bkg$ are obtained from respective fits by Eq.~(\ref{eq:yield}) and Eq.~(\ref{eq:mcos}) as before. 
The obtained signal $\mcos$ can be readily converted to $\rzero$.
This ``manual'' (bin count) calculation of $\rzero$ does not use a signal shape and is therefore not affected by the distortion of the signal shape caused by the momentum and angular smearing. The results are shown in Fig.~\ref{fig:rho}. Again, no qualitative differences are observed from all other methods, presumably because the track-smearing effect is small.

% \blue{ I checked the code, and for the ''manual", we are actually counting bins with pair mass between 1.01 and 1.03 (approximately $m_{\phi}$+- 2 $\Gamma$) and taking the sum of $\mcos$ $\times$ the mass histogram entries minus the $f_{cos, bkg}\times f_{mass,bkg}$ Where $f_{cos, bkg}\times f_{mass,bkg}$ are still from the respective fits of sig+bkg in [1.0,1.04]. 

% So, essentially, we doing the ''default" inv mass fit, and the using bin counting only for the signal and then using the fits for the background in the peak region. Which is approximately the same as [1.0,1.04].

% To be clear, currently, we are NOT fitting the mass distribution and $\mcos$ off-peak, and then using those functions to describe background under the peak. Is that what you want to do?  }

%--------------------------------------------------------------------------------------------------
\section{Summary}
In this work we have investigated the effect of tracking resolutions on $\fm$ spin alignment by toy-model simulations. We examine the yield method and the invariant mass method in extracting the spin alignment observable $\rzero$. Using typical values of tracking angular and momentum resolutions (see Eq.~(\ref{eq:res})), we find the effects on the extracted $\rzero$ from both methods to be small ($<0.0005$), well within typical experimental uncertainties ($\sim 0.001$) on the measured $\rzero$.

\section*{Acknowledgment} 
We are grateful to Dr.~Xin Dong and Dr.~Sergei Voloshin for fruitful discussions. 
This work is supported in part by the U.S.~Department of Energy (Grant No.~DE-SC0012910).

\bibliography{ref}

\begin{thebibliography}{18}%
\makeatletter
\providecommand \@ifxundefined [1]{%
 \@ifx{#1\undefined}
}%
\providecommand \@ifnum [1]{%
 \ifnum #1\expandafter \@firstoftwo
 \else \expandafter \@secondoftwo
 \fi
}%
\providecommand \@ifx [1]{%
 \ifx #1\expandafter \@firstoftwo
 \else \expandafter \@secondoftwo
 \fi
}%
\providecommand \natexlab [1]{#1}%
\providecommand \enquote  [1]{``#1''}%
\providecommand \bibnamefont  [1]{#1}%
\providecommand \bibfnamefont [1]{#1}%
\providecommand \citenamefont [1]{#1}%
\providecommand \href@noop [0]{\@secondoftwo}%
\providecommand \href [0]{\begingroup \@sanitize@url \@href}%
\providecommand \@href[1]{\@@startlink{#1}\@@href}%
\providecommand \@@href[1]{\endgroup#1\@@endlink}%
\providecommand \@sanitize@url [0]{\catcode `\\12\catcode `\$12\catcode
  `\&12\catcode `\#12\catcode `\^12\catcode `\_12\catcode `\%12\relax}%
\providecommand \@@startlink[1]{}%
\providecommand \@@endlink[0]{}%
\providecommand \url  [0]{\begingroup\@sanitize@url \@url }%
\providecommand \@url [1]{\endgroup\@href {#1}{\urlprefix }}%
\providecommand \urlprefix  [0]{URL }%
\providecommand \Eprint [0]{\href }%
\providecommand \doibase [0]{https://doi.org/}%
\providecommand \selectlanguage [0]{\@gobble}%
\providecommand \bibinfo  [0]{\@secondoftwo}%
\providecommand \bibfield  [0]{\@secondoftwo}%
\providecommand \translation [1]{[#1]}%
\providecommand \BibitemOpen [0]{}%
\providecommand \bibitemStop [0]{}%
\providecommand \bibitemNoStop [0]{.\EOS\space}%
\providecommand \EOS [0]{\spacefactor3000\relax}%
\providecommand \BibitemShut  [1]{\csname bibitem#1\endcsname}%
\let\auto@bib@innerbib\@empty
%</preamble>
\bibitem [{\citenamefont {Liang}\ and\ \citenamefont
  {Wang}(2005)}]{Liang:2004ph}%
  \BibitemOpen
  \bibfield  {author} {\bibinfo {author} {\bibfnamefont {Z.-T.}\ \bibnamefont
  {Liang}}\ and\ \bibinfo {author} {\bibfnamefont {X.-N.}\ \bibnamefont
  {Wang}},\ }\bibfield  {title} {\bibinfo {title} {{Globally polarized
  quark-gluon plasma in non-central A+A collisions}},\ }\href
  {https://doi.org/10.1103/PhysRevLett.94.102301} {\bibfield  {journal}
  {\bibinfo  {journal} {Phys. Rev. Lett.}\ }\textbf {\bibinfo {volume} {94}},\
  \bibinfo {pages} {102301} (\bibinfo {year} {2005})},\ \bibinfo {note}
  {[Erratum: Phys.Rev.Lett. 96, 039901 (2006)]},\ \Eprint
  {https://arxiv.org/abs/nucl-th/0410079} {arXiv:nucl-th/0410079} \BibitemShut
  {NoStop}%
\bibitem [{\citenamefont {Voloshin}(2004)}]{Voloshin:2004ha}%
  \BibitemOpen
  \bibfield  {author} {\bibinfo {author} {\bibfnamefont {S.~A.}\ \bibnamefont
  {Voloshin}},\ }\bibfield  {title} {\bibinfo {title} {{Polarized secondary
  particles in unpolarized high energy hadron-hadron collisions?}},\
  }\href@noop {} {\  (\bibinfo {year} {2004})},\ \Eprint
  {https://arxiv.org/abs/nucl-th/0410089} {arXiv:nucl-th/0410089} \BibitemShut
  {NoStop}%
\bibitem [{\citenamefont {Wu}\ \emph {et~al.}(2019)\citenamefont {Wu},
  \citenamefont {Pang}, \citenamefont {Huang},\ and\ \citenamefont
  {Wang}}]{Wu:2019eyi}%
  \BibitemOpen
  \bibfield  {author} {\bibinfo {author} {\bibfnamefont {H.-Z.}\ \bibnamefont
  {Wu}}, \bibinfo {author} {\bibfnamefont {L.-G.}\ \bibnamefont {Pang}},
  \bibinfo {author} {\bibfnamefont {X.-G.}\ \bibnamefont {Huang}},\ and\
  \bibinfo {author} {\bibfnamefont {Q.}~\bibnamefont {Wang}},\ }\bibfield
  {title} {\bibinfo {title} {{Local spin polarization in high energy heavy ion
  collisions}},\ }\href {https://doi.org/10.1103/PhysRevResearch.1.033058}
  {\bibfield  {journal} {\bibinfo  {journal} {Phys. Rev. Research.}\ }\textbf
  {\bibinfo {volume} {1}},\ \bibinfo {pages} {033058} (\bibinfo {year}
  {2019})},\ \Eprint {https://arxiv.org/abs/1906.09385} {arXiv:1906.09385
  [nucl-th]} \BibitemShut {NoStop}%
\bibitem [{\citenamefont {Becattini}\ and\ \citenamefont
  {Lisa}(2020)}]{Becattini:2020ngo}%
  \BibitemOpen
  \bibfield  {author} {\bibinfo {author} {\bibfnamefont {F.}~\bibnamefont
  {Becattini}}\ and\ \bibinfo {author} {\bibfnamefont {M.~A.}\ \bibnamefont
  {Lisa}},\ }\bibfield  {title} {\bibinfo {title} {{Polarization and Vorticity
  in the Quark\textendash{}Gluon Plasma}},\ }\href
  {https://doi.org/10.1146/annurev-nucl-021920-095245} {\bibfield  {journal}
  {\bibinfo  {journal} {Ann. Rev. Nucl. Part. Sci.}\ }\textbf {\bibinfo
  {volume} {70}},\ \bibinfo {pages} {395} (\bibinfo {year} {2020})},\ \Eprint
  {https://arxiv.org/abs/2003.03640} {arXiv:2003.03640 [nucl-ex]} \BibitemShut
  {NoStop}%
\bibitem [{\citenamefont {Sheng}\ \emph {et~al.}(2020)\citenamefont {Sheng},
  \citenamefont {Oliva},\ and\ \citenamefont {Wang}}]{Sheng:2019kmk}%
  \BibitemOpen
  \bibfield  {author} {\bibinfo {author} {\bibfnamefont {X.-L.}\ \bibnamefont
  {Sheng}}, \bibinfo {author} {\bibfnamefont {L.}~\bibnamefont {Oliva}},\ and\
  \bibinfo {author} {\bibfnamefont {Q.}~\bibnamefont {Wang}},\ }\bibfield
  {title} {\bibinfo {title} {{What can we learn from the global spin alignment
  of $\phi$ mesons in heavy-ion collisions?}},\ }\href
  {https://doi.org/10.1103/PhysRevD.101.096005} {\bibfield  {journal} {\bibinfo
   {journal} {Phys. Rev. D}\ }\textbf {\bibinfo {volume} {101}},\ \bibinfo
  {pages} {096005} (\bibinfo {year} {2020})},\ \bibinfo {note} {[Erratum:
  Phys.Rev.D 105, 099903 (2022)]},\ \Eprint {https://arxiv.org/abs/1910.13684}
  {arXiv:1910.13684 [nucl-th]} \BibitemShut {NoStop}%
\bibitem [{\citenamefont {Adamczyk}\ \emph {et~al.}(2017)\citenamefont
  {Adamczyk} \emph {et~al.}}]{STAR:2017ckg}%
  \BibitemOpen
  \bibfield  {author} {\bibinfo {author} {\bibfnamefont {L.}~\bibnamefont
  {Adamczyk}} \emph {et~al.} (\bibinfo {collaboration} {STAR}),\ }\bibfield
  {title} {\bibinfo {title} {{Global $\Lambda$ hyperon polarization in nuclear
  collisions: evidence for the most vortical fluid}},\ }\href
  {https://doi.org/10.1038/nature23004} {\bibfield  {journal} {\bibinfo
  {journal} {Nature}\ }\textbf {\bibinfo {volume} {548}},\ \bibinfo {pages}
  {62} (\bibinfo {year} {2017})},\ \Eprint {https://arxiv.org/abs/1701.06657}
  {arXiv:1701.06657 [nucl-ex]} \BibitemShut {NoStop}%
\bibitem [{\citenamefont {Acharya}\ \emph {et~al.}(2020)\citenamefont {Acharya}
  \emph {et~al.}}]{ALICE:2019aid}%
  \BibitemOpen
  \bibfield  {author} {\bibinfo {author} {\bibfnamefont {S.}~\bibnamefont
  {Acharya}} \emph {et~al.} (\bibinfo {collaboration} {ALICE}),\ }\bibfield
  {title} {\bibinfo {title} {{Evidence of Spin-Orbital Angular Momentum
  Interactions in Relativistic Heavy-Ion Collisions}},\ }\href
  {https://doi.org/10.1103/PhysRevLett.125.012301} {\bibfield  {journal}
  {\bibinfo  {journal} {Phys. Rev. Lett.}\ }\textbf {\bibinfo {volume} {125}},\
  \bibinfo {pages} {012301} (\bibinfo {year} {2020})},\ \Eprint
  {https://arxiv.org/abs/1910.14408} {arXiv:1910.14408 [nucl-ex]} \BibitemShut
  {NoStop}%
\bibitem [{\citenamefont {Abdallah}\ \emph {et~al.}(2023)\citenamefont
  {Abdallah} \emph {et~al.}}]{STAR:2022fan}%
  \BibitemOpen
  \bibfield  {author} {\bibinfo {author} {\bibfnamefont {M.~S.}\ \bibnamefont
  {Abdallah}} \emph {et~al.} (\bibinfo {collaboration} {STAR}),\ }\bibfield
  {title} {\bibinfo {title} {{Pattern of global spin alignment of
  \ensuremath{\phi} and K$^{*0}$ mesons in heavy-ion collisions}},\ }\href
  {https://doi.org/10.1038/s41586-022-05557-5} {\bibfield  {journal} {\bibinfo
  {journal} {Nature}\ }\textbf {\bibinfo {volume} {614}},\ \bibinfo {pages}
  {244} (\bibinfo {year} {2023})},\ \Eprint {https://arxiv.org/abs/2204.02302}
  {arXiv:2204.02302 [hep-ph]} \BibitemShut {NoStop}%
\bibitem [{\citenamefont {Sheng}\ \emph {et~al.}(2023)\citenamefont {Sheng},
  \citenamefont {Oliva}, \citenamefont {Liang}, \citenamefont {Wang},\ and\
  \citenamefont {Wang}}]{Sheng:2022wsy}%
  \BibitemOpen
  \bibfield  {author} {\bibinfo {author} {\bibfnamefont {X.-L.}\ \bibnamefont
  {Sheng}}, \bibinfo {author} {\bibfnamefont {L.}~\bibnamefont {Oliva}},
  \bibinfo {author} {\bibfnamefont {Z.-T.}\ \bibnamefont {Liang}}, \bibinfo
  {author} {\bibfnamefont {Q.}~\bibnamefont {Wang}},\ and\ \bibinfo {author}
  {\bibfnamefont {X.-N.}\ \bibnamefont {Wang}},\ }\bibfield  {title} {\bibinfo
  {title} {{Spin Alignment of Vector Mesons in Heavy-Ion Collisions}},\ }\href
  {https://doi.org/10.1103/PhysRevLett.131.042304} {\bibfield  {journal}
  {\bibinfo  {journal} {Phys. Rev. Lett.}\ }\textbf {\bibinfo {volume} {131}},\
  \bibinfo {pages} {042304} (\bibinfo {year} {2023})},\ \Eprint
  {https://arxiv.org/abs/2205.15689} {arXiv:2205.15689 [nucl-th]} \BibitemShut
  {NoStop}%
\bibitem [{\citenamefont {Workman}\ and\ \citenamefont
  {Others}(2022)}]{Workman:2022ynf}%
  \BibitemOpen
  \bibfield  {author} {\bibinfo {author} {\bibfnamefont {R.~L.}\ \bibnamefont
  {Workman}}\ and\ \bibinfo {author} {\bibnamefont {Others}} (\bibinfo
  {collaboration} {Particle Data Group}),\ }\bibfield  {title} {\bibinfo
  {title} {{Review of Particle Physics}},\ }\href
  {https://doi.org/10.1093/ptep/ptac097} {\bibfield  {journal} {\bibinfo
  {journal} {PTEP}\ }\textbf {\bibinfo {volume} {2022}},\ \bibinfo {pages}
  {083C01} (\bibinfo {year} {2022})},\ \bibinfo {note}
  {https://pdg.lbl.gov/2023/reviews/rpp2022-rev-scalar-mesons.pdf,
  https://pdg.lbl.gov/2023/reviews/rpp2022-rev-non-qqbar-mesons.pdf}\BibitemShut
  {NoStop}%
\bibitem [{\citenamefont {Abelev}\ \emph
  {et~al.}(2009{\natexlab{a}})\citenamefont {Abelev} \emph
  {et~al.}}]{STAR:2008med}%
  \BibitemOpen
  \bibfield  {author} {\bibinfo {author} {\bibfnamefont {B.~I.}\ \bibnamefont
  {Abelev}} \emph {et~al.} (\bibinfo {collaboration} {STAR}),\ }\bibfield
  {title} {\bibinfo {title} {{Systematic Measurements of Identified Particle
  Spectra in $pp$, $d+$Au and Au+Au Collisions from STAR}},\ }\href
  {https://doi.org/10.1103/PhysRevC.79.034909} {\bibfield  {journal} {\bibinfo
  {journal} {Phys. Rev. C}\ }\textbf {\bibinfo {volume} {79}},\ \bibinfo
  {pages} {034909} (\bibinfo {year} {2009}{\natexlab{a}})},\ \Eprint
  {https://arxiv.org/abs/0808.2041} {arXiv:0808.2041 [nucl-ex]} \BibitemShut
  {NoStop}%
\bibitem [{\citenamefont {Anderson}\ \emph {et~al.}(2003)\citenamefont
  {Anderson} \emph {et~al.}}]{Anderson:2003ur}%
  \BibitemOpen
  \bibfield  {author} {\bibinfo {author} {\bibfnamefont {M.}~\bibnamefont
  {Anderson}} \emph {et~al.},\ }\bibfield  {title} {\bibinfo {title} {{The STAR
  time projection chamber: A Unique tool for studying high multiplicity events
  at RHIC}},\ }\href {https://doi.org/10.1016/S0168-9002(02)01964-2} {\bibfield
   {journal} {\bibinfo  {journal} {Nucl. Instrum. Meth. A}\ }\textbf {\bibinfo
  {volume} {499}},\ \bibinfo {pages} {659} (\bibinfo {year} {2003})},\ \Eprint
  {https://arxiv.org/abs/nucl-ex/0301015} {arXiv:nucl-ex/0301015} \BibitemShut
  {NoStop}%
\bibitem [{\citenamefont {Ackermann}\ \emph {et~al.}(2003)\citenamefont
  {Ackermann} \emph {et~al.}}]{STAR:2002eio}%
  \BibitemOpen
  \bibfield  {author} {\bibinfo {author} {\bibfnamefont {K.~H.}\ \bibnamefont
  {Ackermann}} \emph {et~al.} (\bibinfo {collaboration} {STAR}),\ }\bibfield
  {title} {\bibinfo {title} {{STAR detector overview}},\ }\href
  {https://doi.org/10.1016/S0168-9002(02)01960-5} {\bibfield  {journal}
  {\bibinfo  {journal} {Nucl. Instrum. Meth. A}\ }\textbf {\bibinfo {volume}
  {499}},\ \bibinfo {pages} {624} (\bibinfo {year} {2003})}\BibitemShut
  {NoStop}%
\bibitem [{\citenamefont {Abelev}\ \emph {et~al.}(2007)\citenamefont {Abelev}
  \emph {et~al.}}]{STAR:2007mum}%
  \BibitemOpen
  \bibfield  {author} {\bibinfo {author} {\bibfnamefont {B.~I.}\ \bibnamefont
  {Abelev}} \emph {et~al.} (\bibinfo {collaboration} {STAR}),\ }\bibfield
  {title} {\bibinfo {title} {{Partonic flow and phi-meson production in Au + Au
  collisions at $\sqrt{s_{_{\rm NN}}}$ = 200 GeV}},\ }\href
  {https://doi.org/10.1103/PhysRevLett.99.112301} {\bibfield  {journal}
  {\bibinfo  {journal} {Phys. Rev. Lett.}\ }\textbf {\bibinfo {volume} {99}},\
  \bibinfo {pages} {112301} (\bibinfo {year} {2007})},\ \Eprint
  {https://arxiv.org/abs/nucl-ex/0703033} {arXiv:nucl-ex/0703033} \BibitemShut
  {NoStop}%
\bibitem [{\citenamefont {Adams}\ \emph {et~al.}(2005)\citenamefont {Adams}
  \emph {et~al.}}]{STAR:2004jwm}%
  \BibitemOpen
  \bibfield  {author} {\bibinfo {author} {\bibfnamefont {J.}~\bibnamefont
  {Adams}} \emph {et~al.} (\bibinfo {collaboration} {STAR}),\ }\bibfield
  {title} {\bibinfo {title} {{Azimuthal anisotropy in Au+Au collisions at
  $\sqrt{s_{_{\rm NN}}}$ = 200 GeV}},\ }\href
  {https://doi.org/10.1103/PhysRevC.72.014904} {\bibfield  {journal} {\bibinfo
  {journal} {Phys. Rev. C}\ }\textbf {\bibinfo {volume} {72}},\ \bibinfo
  {pages} {014904} (\bibinfo {year} {2005})},\ \Eprint
  {https://arxiv.org/abs/nucl-ex/0409033} {arXiv:nucl-ex/0409033} \BibitemShut
  {NoStop}%
\bibitem [{\citenamefont {Abelev}\ \emph
  {et~al.}(2009{\natexlab{b}})\citenamefont {Abelev} \emph
  {et~al.}}]{Abelev:2008aa}%
  \BibitemOpen
  \bibfield  {author} {\bibinfo {author} {\bibfnamefont {B.~I.}\ \bibnamefont
  {Abelev}} \emph {et~al.} (\bibinfo {collaboration} {STAR}),\ }\bibfield
  {title} {\bibinfo {title} {{Measurements of phi meson production in
  relativistic heavy-ion collisions at RHIC}},\ }\href
  {https://doi.org/10.1103/PhysRevC.79.064903} {\bibfield  {journal} {\bibinfo
  {journal} {Phys. Rev.}\ }\textbf {\bibinfo {volume} {C79}},\ \bibinfo {pages}
  {064903} (\bibinfo {year} {2009}{\natexlab{b}})},\ \Eprint
  {https://arxiv.org/abs/0809.4737} {arXiv:0809.4737 [nucl-ex]} \BibitemShut
  {NoStop}%
%%CITATION = ARXIV:0809.4737;%%
\bibitem [{Note1()}]{Note1}%
  \BibitemOpen
  \bibinfo {note} {One can also subtract the mixed-event background first, and
  then do a fit with BW and a residual background to obtain the signal-to-noise
  ratio with the subtracted mixed-event background added back. Since the
  purpose of this fit is to obtain the signal-to-noise ratio of $\phi {\protect
  \mathrm {\protect \mbox {-}meson}}$, not the yield vs.~$\cos \theta ^*$ as in
  the yield method, the different ways of fit do not cause significant
  systematics}\BibitemShut {NoStop}%
\bibitem [{\citenamefont {Mohanty}\ and\ \citenamefont
  {Xu}(2009)}]{Mohanty:2009tz}%
  \BibitemOpen
  \bibfield  {author} {\bibinfo {author} {\bibfnamefont {B.}~\bibnamefont
  {Mohanty}}\ and\ \bibinfo {author} {\bibfnamefont {N.}~\bibnamefont {Xu}},\
  }\bibfield  {title} {\bibinfo {title} {{Probe the QCD phase diagram with
  phi-mesons in high energy nuclear collisions}},\ }\href
  {https://doi.org/10.1088/0954-3899/36/6/064022} {\bibfield  {journal}
  {\bibinfo  {journal} {J. Phys. G}\ }\textbf {\bibinfo {volume} {36}},\
  \bibinfo {pages} {064022} (\bibinfo {year} {2009})},\ \Eprint
  {https://arxiv.org/abs/0901.0313} {arXiv:0901.0313 [nucl-ex]} \BibitemShut
  {NoStop}%
\end{thebibliography}%
\end{document}